\documentclass[doublecol]{epl2} 
\usepackage{graphicx}
\usepackage{dcolumn}
\usepackage{bm}
\usepackage{environ}

\title{Strain Enhanced Superconductivity in Li-Doped Graphene}
\shorttitle{Strain Enhanced Superconductivity in Li-Doped Graphene}
\author{Jelena Pe\v{s}i\'{c}\inst{1}, Rado\v{s} Gaji\'{c}\inst{1}, Kurt Hingerl\inst{2}\and Milivoj Beli\'{c}\inst{3}}

\shortauthor{Jelena Pe\v{s}i\'{c}\etal}

\institute{
  \inst{1} Institute of Physics, University of Belgrade, Pregrevica 118, 11080 Belgrade, Serbia\\
  \inst{2}Center for Surface and Nanoanalytics, Johannes Kepler University, Linz, Austria\\
 \inst{3} Science Program, Texas A\&M University at Qatar, P.O. Box 23874 Doha, Qatar}


\abstract{
We present a new way to enhance the electron-phonon coupling constant and the critical superconducting temperature of graphene, significantly beyond all reported values. Using density functional theory, we explore the application effects of the tensile biaxial strain on the lithium intercalated graphene. Both effects together, the presence of adatom and the strain, trigger enhancement of critical temperature, up to 300\%, compared to non-strained lithium intercalated graphene.}

\begin{document}

\maketitle

\section{INTRODUCTION}

Graphene, a single atomic layer of carbon atoms arranged in a honeycomb lattice, has been attracting remarkable attention for its unique properties ever since it was successfully isolated in 2004\cite{Nov}. As the first case of real-2D atomic crystals (MoS$_2$, BN, NbSe$_2$, Bi$_2$Te$_3$, InSe,…) graphene proved to be an interesting system with properties essentially different from 3D objects.  In addition, due to a hexagonal lattice with two carbon atoms per unit cell, the electronic spectrum of graphene has a Dirac cone analog to relativistic massless fermions. As a consequence, the chiral Dirac electrons appear, resulting in peculiar quantum effects like Klein tunneling \cite{Katsnelson}, Zitterbewegung of electrons \cite{Katsnelson2, Rusin} or anomalous quantum Hall effect \cite{Zhang, Nov2}. On the other hand, the electron-phonon interaction in graphene is also quite interesting. Namely, because of the specific Fermi surface, there is a pronounced Kohn anomaly like in graphite \cite{Piscanec} around $\Gamma$ and K points \cite{Zhou}. In addition, the peculiar electronic band structure of graphene gives rise to the breakdown of the adiabatic Born-Oppenheimer approximation \cite{Pisana}. Extraordinary effects in graphene make it a promising material for future research as well as for various applications. Nevertheless, one important effect has been missing on graphene list of the exceptional properties so far. This is superconductivity that has not been observed yet neither in pristine nor doped graphene although considerable theoretical efforts have been invested in exploring possible pairing instabilities\cite{EZhao,RoyHerbut, Honerkamp}.

Particularly, the existence of superconductivity in the graphene was highly investigated subject in the past few years. Various superconducting mechanisms have been extensively researched. Numerous superconducting models in graphene have been proposed e.g. a chiral superconductivity \cite{Nandkishore}, an interlayer pairing of chiral electrons \cite{Hosseini} as well as many other exotic pairing mechanisms \cite{Roy} that have been studied \cite{Milica}.
Whereas all these theoretical studies result in the possible appearance of the superconductivity in graphene, the electron-phonon mechanism is still the most probable cause for potential superconductivity. Nonetheless, this is the naturally first choice in the search for superconductivity in 2D carbon layer, pristine as well as intercalated one \cite{Einenkel, Profeta, Lozovik, Mazin, Dietel}. Moreover, the phonon mediated superconductivity has been observed and proven by photoemission spectroscopy \cite{Valla} in graphite intercalated compounds \cite{Dresselhaus, Jishi, Emery, Calandra2005}.

 Strong electron-phonon coupling could be caused by a large number of carriers, strong deformation potentials, and coupling of electrons and phonon modes at low energies. Unfortunately pristine graphene does not fulfil any of these. The enhancement of coupling to the out-of-plane vibrations is an imperative for appearance of superconductivity so the new states at the Fermi level should be introduced, in the manner similar to the graphite intercalated compound (GIC). Intercalation of atoms into the layered materials has been known as a method of introduction of new properties in the pristine materials \cite{Attaccalite}.

\begin{figure}
\onefigure{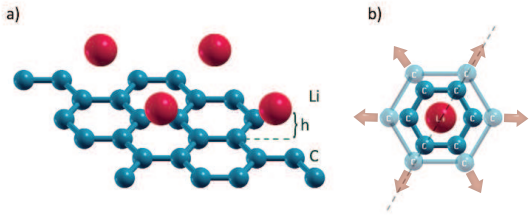}
\caption{\label{fig1}(Color online) a) Lithium intercalated graphene, h is adatom graphene distance. b) Schematic description of biaxial tensile strain }
\end{figure}
GIC doped with various alkaline atoms displays superconductivity in many cases\cite{Dresselhaus, Jishi, Emery}. In the superconducting GIC at the Fermi level, an interlayer band has been formed and it enhances the electron-phonon coupling constant $\lambda$ \cite{Dresselhaus}. Therefore, due to the presence of interlayer band, the number of carriers is gets augmented, the coupling to carbon out-of-plane vibrations is enhanced and there is coupling to the intercalant vibrations as well. The electron-phonon coupling constant is proportional to the DOS at the Fermi level and the deformational potential D and inversely proportional to the square of phonon frequency of the mode coupled to electrons (1)
\begin{eqnarray}
\lambda=\frac{N(0)D^2}{M\ {\omega }^2_{ph}}\
\end{eqnarray}
In equation (1) ${\omega }_{ph}$ is phonon frequency of phonon vibrations involved in superconductivity, N(0) is electronic DOS per spin at the Fermi level and M is an effective atomic mass. For the appearance of superconductivity it is necessary to have the charge transfer to graphene layer, but also the interlayer band must be formed on the Fermi level. The electron-phonon coupling constant increases with the deformation potential D which depends on the distance between adatoms and the graphene, $h$. The smaller the distance is, the larger D becomes. The distance $h$ cannot be decreased infinitely, since a too small $h$ will cause a complete charge transfer and an upshift of the intercalant band, which then becomes empty and forms above the Fermi level. Another important prerequisite for the introduction of superconductivity is the existence of coupling between carbon out-of-plane vibrations and electrons (which is essential since the coupling with in-plane vibration is not large enough and $\lambda$ has small value). This is achieved by the transitions between the graphene $\pi$* states and the interlayer band.

The different intercalants and geometries allow tuning of the properties of carbon, producing the largest increase in $T_c$ (Calcium doped graphite 11.5K) \cite{Weller, Emery}. Although GICs show interesting properties, not all types of intercalant atoms produce superconductivity or significantly increase the $T_c$. It seems that the charge transfer from the interlayer is crucial. Though charge transfer is necessary, the completion of the charge transfer is deleterious for the enhancement of the superconductivity. In some of GICs, the charge transfer between adatom and graphene is incomplete and they display a superconductivity, on the other hand, like Li-GIC, where the charge transfer is complete resulting in low $T_c$ (0.9K). In the Li-GIC, a strong confinement for electrons along the z-axis exists and it prevents occupation of the interlayer state. Because the quantum confinement is removed in monolayer \cite{Calandra, Profeta} this results in reduction of charge transfer and it is beneficial for superconductivity. Thus the Li doped graphene is shown to be superconductive (8.1K) with much higher $T_c$ than in Ca-doped graphene (1.4K).
After the superconductivity in Li intercalated graphene has been established, much attention has been devoted to study of its enhancement \cite{Kaloni, Szcz, Guzman}. As a conclusion beside to the Li doping, an increase of electron-phonon paring potential is necessary. In order to increase $\lambda$, the phonon frequencies must be softened.

Based on this concept, we study the effects of the tensile equibiaxial strain on the Li-intercalated graphene. Application of the strain is an intensively studied topic, both on theory and experiment \cite{Hua, Huangg, Pereiraa, Lee}. Namely, an application of strain on graphene can induce changes of the vibrational properties \cite{Mohiuddin, Ding}, in the electronic band gaps \cite{Guinea, Levy} and significant changes in conductivity both local and macroscopic level\cite{Teague, Huang, Fu}.
The type of the strain is a very important feature, since the graphene$'$s lattice symmetry determines its band structure.  Breaking of the hexagonal symmetry will modify the band structure of graphene \cite{Zhou, Xiao}, causing the opening of the band gap and many other effects \cite{Marianetti, gui}. Since our intention is to soften modes, without drastically modifying the structure, the tensile equibiaxial strain is employed in the calculations on the Lithium intercalated graphene (LiG) (Fig1 b). Here it is shown that such a strain causes softening of the phonons, in particular,the in-plane phonons will be dramatically softened, whereas the out-of-plane ones will be less affected \cite{Marianetti}. This greatly affects the $\lambda$.

We investigate the enhancement of the electron-phonon interaction in LIG using the first-principle density functional theory (DFT) calculation in the local density approximation (LDA) and based on the prior discussion, we find $\lambda$ is sensitive to the tensile equibiaxial strain, therefore producing a higher $T_c$. For instance, the strain of 10\% makes a T$_c$ increase of almost 300\%!

\section{COMPUTATIONAL DETAILS}

As mentioned above, we employed DFT with LDA \cite{PZ}, using Quantum Espresso (version 5.0.3)\cite{QE}. The ionic positions in the cell are fully relaxed, in all calculations, to their minimum energy configuration using the Broyden-Fletcher-Goldfarb-Shanno (BFGS) algorithm. The hexagonal cell parameter $c$ was set to c=12.5\AA \ in order to simulate a two-dimensional system and avoid an interaction due to periodicity. The norm-conserving pseudopotential and the plane wave cutoff energy of 65 Ry were used in the calculation. Although DFT with LDA may have problems in application in certain situations where electronic correlations are strong, for graphene, with large electronic bands, it is quite a suitable assumption\cite{Kotliar, Marianetti}.
As stated before, there is the Kohn anomaly. Although DFT is known to underestimate the electron-exchange correlation energy in the presence of the Kohn anomaly\cite{Lazzeri}, the usage of DFT here is justified. The differences appear only in a small portion of the first Brilluen zone and do not lead to significant inconsistencies when the electron interaction with entire phonon system is observed \cite{Borysenko}.
The unit cell for the LiG monolayer was modeled in the $\sqrt{3}$X$\sqrt{3}$R $60^{\circ}$ in-plane unit cell, consisting of one Li atom placed above the center of the carbon hexagon, with an adatom-graphene distance $h=1.8$ \AA. (Figure 1). $\lambda$ was calculated with the electron momentum k-mesh up to 40x40x1 and the phonon q-mesh 20x20x1.
The Eliashberg function defined as:

\begin{eqnarray*}
{\alpha }^2F\left(\omega \right)=\frac{1}{N(0)N_kN_q}\sum_{n{\mathbf k},\ m{\mathbf q},\nu }{{\left|g^{\nu }_{n{\mathbf k},\ mk+{\mathbf q}}\right|}^2}\times\\ \delta ({\varepsilon }_{n{\mathbf k}})\delta ({\varepsilon }_{m{\mathbf k}{\mathbf +}{\mathbf q}})\delta (\omega -{\omega }^{\nu }_{{\mathbf q}})
\end{eqnarray*}

where $N(0)$ is the total DOS per spin and $N_k$ and $N_q$ the total numbers of $k$ and $q$ points, respectively. The electron eigenvalues are labelled with the band index ($n$ and $m$) and the wavevector ($k$ and $k+q$), while the phonon frequencies with the mode number ($\nu$) and the wavevector ($q$). ${g^{\nu }_{n{\mathbf k},\ mk+{\mathbf q}}}$ represents the electron-phonon matrix element. The total electron-phonon coupling $\lambda$($\omega$) is given:
\begin{eqnarray}
\lambda \left(\omega \right)=2\int^{\omega }_0{d\omega '\frac{{\alpha }^2F(\omega ')}{\omega '}}
\end{eqnarray}
The superconducting critical temperature was estimated using the Allen-Dynes formula with $\mu$*=0.112 \cite{Allen}
\begin{eqnarray}
T_c=\frac{{\omega{}}_{log}}{1.2}exp\left[\frac{-1.04(1+\lambda{})}{\lambda{}\left(1-0.62{\mu{}}^*\right)-{\mu{}}^*}\right]
\end{eqnarray}
Where
\begin{eqnarray}
{\omega{}}_{log}=exp\left[\frac{2}{\lambda{}}\int\frac{d\omega{}}{\omega{}}{\alpha{}}^2F(\omega{})log\omega{}\right]
\end{eqnarray}

\section{RESULTS}
In order to strain LIG monolayer and increase the lattice constant, the in-plane distance between C atoms, is increased leaving the hexagonal symmetry preserved. The Li adatom is placed above the H site in graphene (the center of hexagon) (Figure 1), which, according to DFT study is the favorable adsorption site \cite{Chan}. The modification of the lattice constant does not interfere with the Li adatom position which remains fixed in the center of the hexagon, leaving the symmetry unbroken. Due to the expansion of the carbon atom distances and the invariance of the hexagonal symmetry, the Li adatom shifts only along the z axis. The effects of several values of the strain, which increase the lattice constant by 3\%, 5\%, 7\%, and 10\% , are studied. Larger strains are not applied due to the instabilities that occur after the attempt of geometrical optimization and relaxation.
\begin{table*}[t]
\caption{Physical properties of graphene under different values of tensile
equibiaxial strain}
\begin{tabular}{p{39pt}p{102pt}p{69pt}p{63pt}p{53pt}p{56pt}}

\parbox{39pt}{\centering
\textit{Strain \%}
} & \parbox{102pt}{\centering
\textit{ h distance (\AA{})}
} & \parbox{69pt}{\centering
\textit{C-C bond lenght (\AA{})}
} & \parbox{63pt}{\centering
\textit{$\lambda{}$}
} & \parbox{53pt}{\centering
\textit{$\omega{}$$_{log}$}
} & \parbox{56pt}{\centering
\textit{$T_c$}
} \\
\hline
\parbox{39pt}{\centering
0\%
} & \parbox{102pt}{\centering
1.80
} & \parbox{69pt}{\centering
1.42
} & \parbox{63pt}{\centering
0.61
} & \parbox{53pt}{\centering
278.88
} & \parbox{56pt}{\centering
8.1
} \\
\hline
\parbox{39pt}{\centering
3\%
} & \parbox{102pt}{\centering
1.69
} & \parbox{69pt}{\centering
1.46
} & \parbox{63pt}{\centering
0.47
} & \parbox{53pt}{\centering
876.17
} & \parbox{56pt}{\centering
6.42
} \\
\hline
\parbox{39pt}{\centering
5\%
} & \parbox{102pt}{\centering
1.64
} & \parbox{69pt}{\centering
1.49
} & \parbox{63pt}{\centering
0.49
} & \parbox{53pt}{\centering
976.20
} & \parbox{56pt}{\centering
9.43
} \\
\hline
\parbox{39pt}{\centering
7\%
} & \parbox{102pt}{\centering
1.61
} & \parbox{69pt}{\centering
1.52
} & \parbox{63pt}{\centering
0.55
} & \parbox{53pt}{\centering
1009.05
} & \parbox{56pt}{\centering
14.73
} \\
\hline
\parbox{39pt}{\centering
10\%
} & \parbox{102pt}{\centering
1.54
} & \parbox{69pt}{\centering
1.57
} & \parbox{63pt}{\centering
0.73
} & \parbox{53pt}{\centering
827.09
} & \parbox{56pt}{\centering
28.72
} \\

\end{tabular}
\end{table*}

Table 1 presents the physical parameters of the Li doped graphene under the various strains. The distance between the Li adatom and graphene decreases with the strain, as the Li adatom moves down deeper towards graphene. When strain is applied, the distance between neighboring C atoms increases and the graphene $\pi$ bonds less repulse Li adatom, which then moves down along the z axis.
\begin{figure}
\onefigure{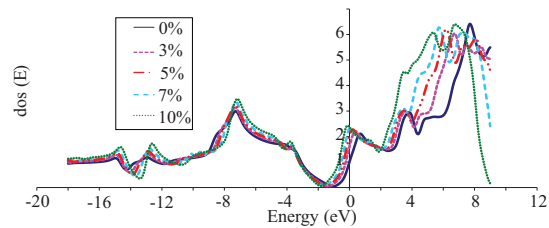}
\caption{\label{fig2}(Color online) Electron density of states for LiC6 under tensile equibiaxial strain }
\end{figure}
In Figure 2, the electron DOS is shown. The small shift of the Fermi level can be observed with the strain. Here is worth mentioning that in graphene, a truly 2D system with low electron density, the long-range Coulomb force is weakly screened and the electron-electron interaction cannot be neglected. The two-dimensionality in graphene can cause enhanced excitonic effects, like the M-point exciton\cite{Alexander, Alexander2} or the charge density waves (CDW) formation (the Peierls transition)  as a result of the Fermi nesting. Especially an interplay between superconductivity and CDW seems to be important \cite{Silva, Comin}.Namely, CDW makes a pre-existing
environment for superconductivity \cite{Chang}.
 Various strains in graphene have been studied as a method for introduction of different broken symmetry phases. CDWs in graphene have been thoroughly investigated both theoretically \cite{Kotov, Khveshchenko, BRoy, Toke} and experimentally \cite{Rahnejat} showing the interesting results. For instance, the presence of axial magnetic field, caused by a buckle strain can leads to realization of CDW\cite{BRoy}. Also, in CaC$_6$ electron-electron repulsion is dominant within graphene sheets\cite{Rahnejat} producing the CDW stripes. In the case of strained LIG, DOS near the Dirac point gets enhanced (Fig 2) hence the question about interplay of electron-electron and electron-phonon interaction can be imposed. Problem of CDW in graphene, doped and strained, is discussed comprehensively and its very existence in LIG is not in conflict with our discussion and results. Moreover, CDW and superconductivity appear together in different systems like high T$_c$ superconductors or intercalated graphite \cite{Silva, Comin, Chang} and can be even used as a criterion for high temperature superconductivity.

 Considerable changes are present in the phonons.
\begin{figure}
\onefigure{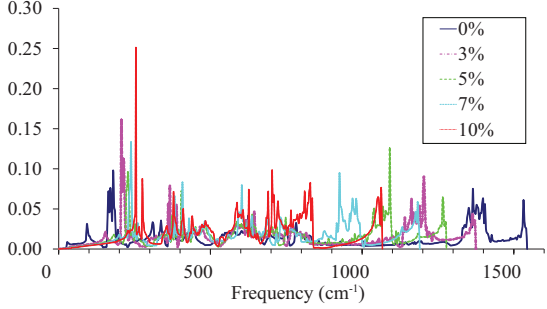}
\caption{\label{fig3}(Color online) Comparison of phonon DOS for various strain }
\end{figure}
In the phonon dispersion spectrum three regions can be distinguished: the adatom-related modes are associated with low energy regions (0-400cm$^{-1}$), where 300-400 cm$^{-1}$ are Li modes mixed with the out-of-plane carbon modes ($C_z$), the mid region (400-900cm$^{-1}$) can be associated with $C_z$ modes and the high energy region with carbon-carbon stretching modes \cite{Profeta}.  The main contributions to $\lambda$ come from the low-energy lithium modes and the carbon vibrations along the z axis, with an additional contribution from the C-C stretching modes (in agreement with \cite{Profeta} and \cite{Calandra}).

Phonon Density of States (PDOS) as a function of strain is depicted in Figure 3. Although low energy modes slightly move upwards in energy, the main effect on the electron-phonon coupling is the softening of graphene high energy C-C stretching modes. They significantly soften with the strain. In contrast to non-strained LIG, the stretching modes have a main influence on $\lambda$ (Figure 4). The Eliashberg spectral function (Figure 4) describes which phonon modes couple with the electrons on the Fermi level. The intensity of the Eliashberg function is greatly increased in the area of the C-C stretching modes, with the strain. This results in a great increase of $\lambda$ and $T_c$. For 10\% tensile equibiaxial strain we get $\lambda$=0.73 and $T_c$=29K.
$\lambda$ is presented as function of strain in Figure 5. It is worth mentioning there is a reduction in $\lambda$ for small values of strain (0-3\%) (Figure 5).
\begin{figure}
\onefigure{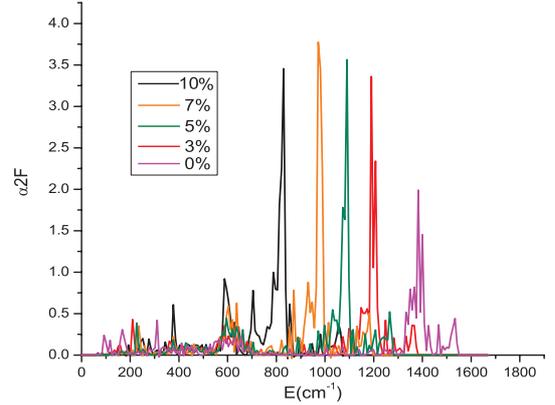}
\caption{\label{fig4}(Color online) Comparison of Eliashberg function for equibiaxial strain }
\end{figure}
Particularly, with strain, the C-C bonds expand, causing a decrease of the Coulomb repulsion between the $\pi$ orbitals and the Li adatom. That allows the Li adatom to come down toward the center on the graphene hexagon. As emphasized before, a too small intercalant-graphite layer distance in the GIC is destructive for superconductivity.  On the other hand, this effect vanishes for larger strains, while an increase in T$_c$, even up to three times larger than the value reported for non-strained LIG, can be observed. This effect is associated to an overlap of the carbon $\pi$ and the Li orbitals.
\begin{figure}
\onefigure{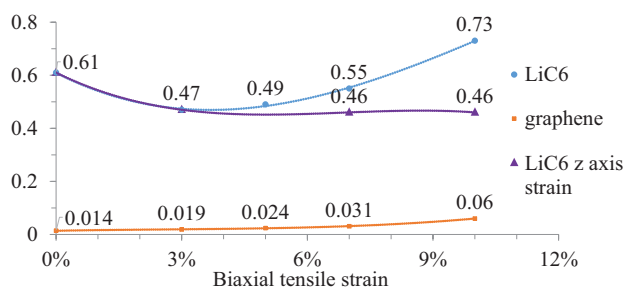}
\caption{\label{fig5}(Color online) Electron phonon coupling constant behavior with tensile equibiaxial strain }
\end{figure}

For small strain, the Li adatom drops down toward the center of hexagon and its orbitals overlap more with the carbon $\pi$ orbitals. That causes an increase in charge transfer and emptying of the interlayer band, which reduces $\lambda$. When more strain is applied, the carbon bonds are elongated and the $\pi$ orbitals move away, both from each other and the center of the hexagon. The orbital overlap is reduced, and after the certain critical value, $\lambda$ increases, following the strain (circle in Figure 5). In order to corroborate this interpretation, we perform two additional calculations: the calculation on non-strained graphene, where the Li adatom position is shifted along z axis; and the second one with the strained pristine graphene. Here it is proven that an increase of $\lambda$ is a mutual effect of strain and doping. For the first calculation, the $\pi$ orbitals remain fixed in their positions (since there is no strain). The overlap with the Li and $\pi$ orbitals increases and one can clearly see that $\lambda$ is decreased (violet triangles in Figure 5) due to an approach to the charge transfer completion and emptying of the interlayer band. The effects of the strain on the pristine graphene$'$s $\lambda$ are also depicted in Figure 5 (orange squares). Graphene has a very small $\lambda$ which isa increased with strain almost four times, but effect of this enhancement is negligible ($\lambda$=0.06).

\begin{figure}
\onefigure{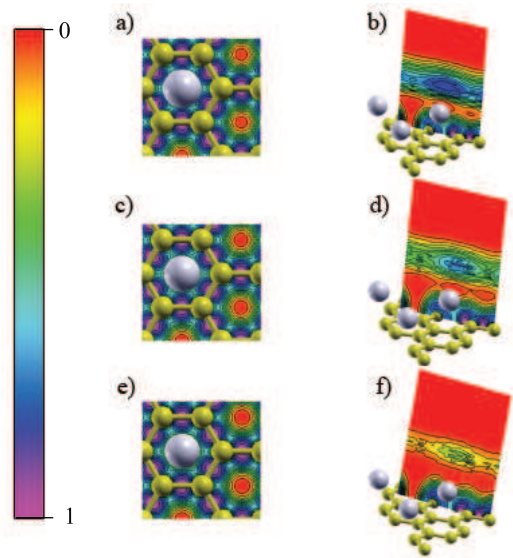}
\caption{\label{fig6}(Color online) ELF (electron localization function) for LIG with strain a) ELF for LIG without strain xy plane b) ELF for LIG without strain xz plane c) ELF for LIG for 5\% strain xy plane d) ELF for LIG for 5\% strain xz plane. e) ELF for LIG for 10\% strain xy plane f) ELF for LIG for 10\% strain xz plane. In a, c and e figures we can see slight changes in ELF projected on xy plane, localization region at Li adatom is enlarged. In b, d and f figures are shown effects of strain, projected on xz plane. Notable change is present for 10\% of strain, where electron localization region is significantly lowered due to described effects.  }
\end{figure}

On the other hand Figure 6, presents the effects of the different strain on electronic localization function (ELF). The significant changes for the large strain are presented, proving the above described effects. For ELF at 10\% of the strain, the electron localization region is greatly lowered as graphene and adatom separate one from another and as a C-C bond are elongated.\\
As expected, the strain alone will not boost the $\lambda$ considerably, nor the doping itself. A complex mechanism of the enhancement is a mutual effect of the mechanical effects with the presence of the interlayer level, all owing to the unique structure of graphene. For the notable enhancement of $\lambda$ a presence of both the adatom and the strain is essential.

\section{CONCLUSIONS}
In this work, using DFT, we have studied the effects of tensile equibiaxial strain on the enhancement of $\lambda$ in the LIG. Since no symmetry is broken, there are no major changes in the electronic structure of the system. On the other hand, strain softens the phonon modes significantly. The critical temperature is enhanced by the strain, up to $T_c$=29K where the electron-phonon coupling constant is 0.73.  We conclude that both the presence of the adatom and the strain is necessary for the enhancement of the $\lambda$.
It is important to stress that this increase in $T_c$, achieved by the described mechanism, can be experimentally realized. A pristine graphene is experimentally confirmed to be elastically stretchable up to 25\% \cite{Lee}. The recent study confirms a fabrication of the intercalated graphene \cite{Yang} which additionally raises an interest in such compounds and their properties. Therefore, an experimental realization of the high $T_c$ superconducting intercalated graphene is to be anticipated, opening a completely new field of the graphene applications.

\acknowledgments
DFT calculations are performed using computational resources at Johannes Kepler University, Linz, Austria. This work was supported by the Serbian Ministry of Education, Science and Technological Development under projects OI 171005. This research is also supported by the project NPRP 7-665-1-125 from the Qatar National Research Fund (a member of the Qatar Foundation).


\end{document}